\documentclass{article}

\usepackage{arxiv}

\usepackage[utf8]{inputenc} 
\usepackage[T1]{fontenc}    
\usepackage{hyperref}       
\usepackage{url}            
\usepackage{booktabs}       
\usepackage{amsfonts}       
\usepackage{nicefrac}       
\usepackage{microtype}      
\usepackage{lipsum}		
\usepackage{graphicx}
\usepackage{natbib}
\usepackage{doi}
\usepackage{multicol}

\title{SceML - A Graphical Modeling Framework for Scenario-based Testing of Autonomous Vehicles}

\author{Barbara Schütt\\
	FZI Research Center for Information Technology\\
	\texttt{schuett@fzi.de} \\
	\And
	Thilo Braun\\
	FZI Research Center for Information Technology\\
	\texttt{braun@fzi.de} \\
	\And
	Stefan Otten\\
	FZI Research Center for Information Technology\\
	\texttt{otten@fzi.de} \\
	\And
	Eric Sax\\
	FZI Research Center for Information Technology\\
	\texttt{sax@fzi.de} \\
}

\renewcommand{\shorttitle}{SceML Modeling Framework}

\hypersetup{
	pdftitle={A template for the arxiv style},
	pdfsubject={q-bio.NC, q-bio.QM},
	pdfauthor={David S.~Hippocampus, Elias D.~Striatum},
	pdfkeywords={First keyword, Second keyword, More},
}

\begin{document}
	\maketitle

	\begin{abstract}
		Ensuring the functional correctness and safety of autonomous vehicles is a major challenge for the automotive industry.
		However, exhaustive physical test drives are not feasible, as billions of driven kilometers would be required to obtain reliable results.
		Scenario-based testing is an approach to tackle this problem and reduce necessary test drives by replacing driven kilometers with simulations of relevant or interesting scenarios.
		These scenarios can be generated or extracted from recorded data with machine learning algorithms or created by experts. 
		In this paper, we propose a novel graphical scenario modeling language.
		The graphical framework allows experts to create new scenarios or review ones designed by other experts or generated by machine learning algorithms.
		The scenario description is modeled as a graph and based on behavior trees.
		It supports different abstraction levels of scenario description during software and test development.
		Additionally, the graph-based structure provides modularity and reusable sub-scenarios, an important use case in scenario modeling.
		A graphical visualization of the scenario enhances comprehensibility for different users. 
		The presented approach eases the scenario creation process and increases the usage of scenarios within development and testing processes.
		
	\end{abstract}

	
	\section{Introduction}
	
	\begin{figure*}[ht]
		\centering
		\includegraphics[width=\textwidth]{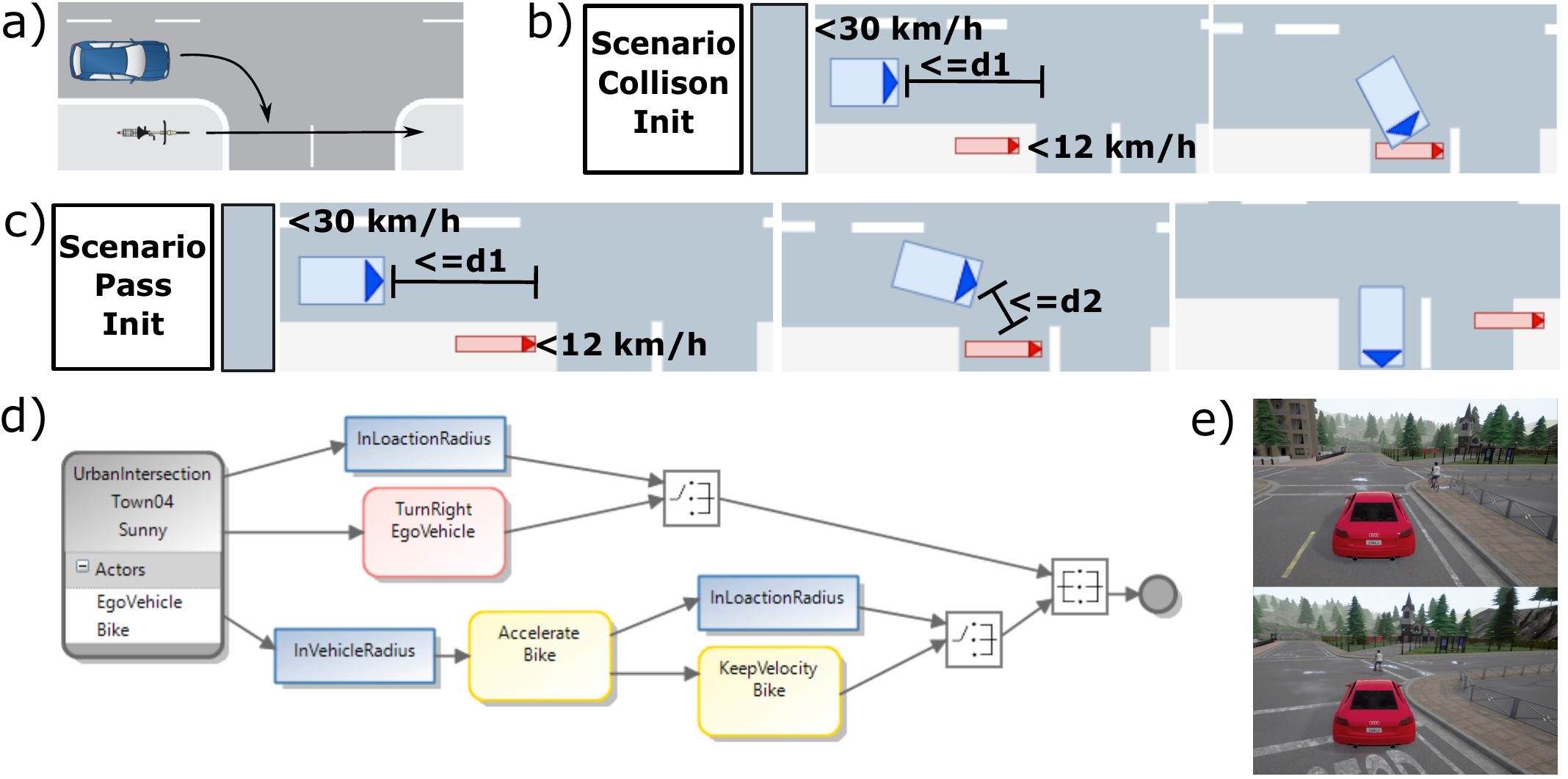}
		\caption{Different presentations of the urban intersection scenario: a) top view of scenario start situation; b) Traffic Sequence Charts (TSL) of the scenario resulting in a collision; c) TSL of the scenario without collision; d) SceML representation; e) simulation results in the CARLA simulator.}
		\label{fig:complete}
	\end{figure*}
	
	In software development, testing is an important step to realize develop systems.
	In particular, in the automotive industry software validation and verification cannot be neglected and constitutes an integral part of the development process to make sure requirements are met and the system fulfills the intended use cases \cite[p.~76]{wood_safety_2019}. 
	According to \citet{wachenfeld2016release}, real world test drives, known as distance-based testing, are a valid testing method and give the most reliable results. 
	However, distance-based testing is not feasible: 
	Almost 8 billion kilometers are necessary to make a statistically significant comparison at a confidence of 95\% to show the failure rate of an autonomous vehicle is lower than the human driver failure rate. 
	This results in 225 years of driving for a fleet of 100 vehicles and if there are changes or variations in the software, all the driving has to be repeated \cite{kalra2016driving,wachenfeld2016release}.
	
	An alternative is offered by scenario-based testing. 
	Compared to the random test cases emerging during distance-based testing, scenario-based testing is a systematic approach to find new test cases \cite{6layer}.
	The goal is to set up a collection of validated and relevant scenarios, depending on the test subject. 
	This approach can be used at different stages during the development process \cite{menzel2018scenarios}. 
	Scenarios are developed at an abstract level during the concept phase, e.g. functional scenarios as shown in Fig.~\ref{fig:abstractionlevel}, and get more and more refined throughout the software development and testing process.
	Distance-based testing is a statistical verification approach, whereas scenario-based testing offers a more direct way to find and examine relevant or critical traffic situations.
	But unlike formal verification it cannot claim completeness of the input space over all scenarios.
	
	\citet{wiecher2019test} proposed an example for the usage of scenarios within the development process, called Test-Driven Scenario Specification, where new requirements specification and analysis techniques are based on executable scenarios and automated testing.
	But in general, scenarios for testing have to be painstakingly created or collected. 
	There are various methods for scenario generation or extraction, e.g. generating test scenarios with machine learning \cite{abdessalem_testing_2018}, combinatorial test generation with constraints \cite{rocklage_automated_2017}, extracting scenarios from test drives \citep{Pfeffer2019,Hartjen2019} or expert-based scenario creation via one of several simulation tools that provide a scenario editor \cite{ipg,dspace,vtd}.
	An appropriate scenario description and modeling language is needed during the scenario acquisition and software development and test process, but widely supported standards are just about to emerge.
	
	Here, we propose a scenario modeling language with the following features:
	\begin{itemize}
		\item It has a graphical scenario modeling concept suitable for human experts.
		\item A scenario is modeled with a graph structure, that can be easily converted in a machine readable data structure.
		\item It allows different depth of information modeling to support different abstraction levels during the development process \cite{menzel2018scenarios}.
		\item It supports modularity in form of sub-modules that can be reused in other scenarios or collected in module data bases or libraries.
		\item The scenario outcome is not defined in the scenario description, since it might not be known before execution and can also be defined in pass and fail criteria or evaluation metrics of a test case where the scenario is used. 
	\end{itemize}
	
	Throughout this work, the following common urban scenario is used to explain state-of-the-art methods and our own approach:
	The ego vehicle, the autonomous vehicle to be tested, and a bike start at an intersection. 
	The bike plans to cross the road, whereas the ego vehicle wants to take a right turn.
	
	In Section \ref{sec:scenario}, a scenario terminology and design for scenario-based testing of automated and autonomous vehicles is described.
	Section \ref{sec:language} introduces our graph-based modeling language notation, our metamodel and an exemplary realization of running our modeled example scenario before Section \ref{sec:conclusion} gives a short conclusion.

	\section{Related Work}
	
	\label{sec:scenario}
	
	\subsection{Scenario Abstraction Levels}
	\begin{figure}[tb]
		\centering
		\includegraphics[width=0.95\linewidth]{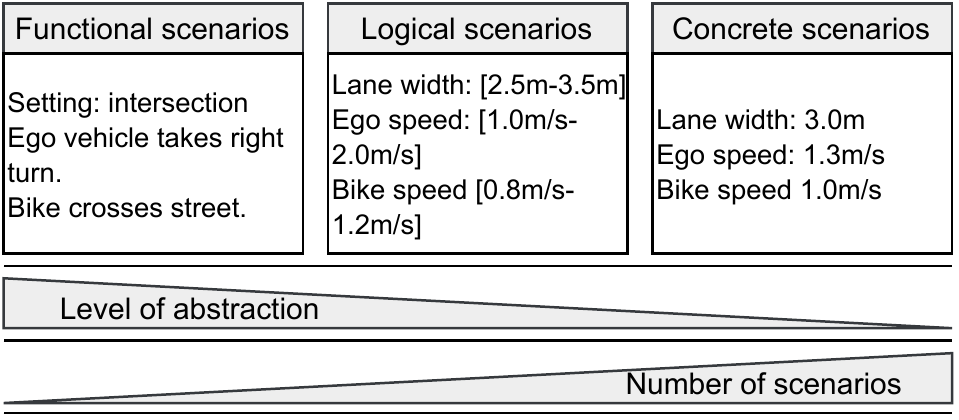}
		\caption{All three abstraction levels as proposed by \cite{menzel2018scenarios}.}
		\label{fig:abstractionlevel}
	\end{figure}
	\citet{menzel2018scenarios} suggest three abstraction levels for scenarios that can accompany the development process and be converted into each other.
	Fig.~\ref{fig:abstractionlevel} illustrates these three levels with the help of our intersection scenario:
	
	The most abstract level of scenario representation are called \textit{functional} scenarios and described via linguistic scenario notation using natural, non-structured language terminology. 
	The main goal for this level is to create scenarios easily understandable for experts. 
	It describes the base road network, in our case the urban intersection map, and all actors with their maneuvers, right turning ego vehicle and road crossing cyclist.
	
	The next abstraction level are \textit{logical} scenarios and refine the representation of functional scenarios with the help of parameters. These parameters can, for instance, be ranges for road width, vehicle positions and velocity or time and weather variables. 
	The parameters can either be described as parameter ranges or with a probability distribution. 
	The last level are concrete scenarios and depict operating scenarios with concrete values for each parameter in the parameter space. 
	This means, that one logical scenario can yield a large number of concrete scenarios, depending on the number of variables, size of range and step size for these ranges.
	
	\subsection{Scenario Description Methods and Languages}
	Motivated from the many different use cases of scenarios, there exist several methods and languages to describe scenarios. 
	One important trade-off is between intuitive usage of a language that is human readable and a well-defined machine-readable executable language that can be used for simulation tools. 
	Graphical description and formalized scenario description languages are the two main approaches in describing scenarios.
	A common approach to describe a traffic scenario between experts is a 2-dimensional view as shown in Fig.~\ref{fig:complete}a) \cite{rosenberger2019towards,wachenfeld2016worst,yasar2008computational}.
	This picture conveys an idea of what is supposed to happen, but lacks crucial information, e.g., which actor is going first, synchronization points or conditional behavior.
	\citet{autovoyage} further developed this concept to a combination of this 2-dimensional view and a natural language description of the scenario and its result.
	\citet{damm2017traffic} proposed Traffic Sequence Charts (TSL), which are depicted in Fig.~\ref{fig:complete}b) and c). 
	This method solves the missing information problem and also defines a powerful graphical notation in describing a scenario and an underlying first-order logic.
	A scenario consists of several snapshots with information about each important event ordered as a sequence.
	However, it defines the scenario outcome which might not be known before a simulation was executed.
	Therefore, both outcomes have to be modeled as scenarios in TSL, shown in Fig.~\ref{fig:complete}b) and c).
	Also, the number of possible outcomes grows in case a scenario contains more than ego vehicle and one actor, since the ego vehicle has more collision opportunities during execution.
	
	The outcome is influenced by scenario variables, e.g. actors' position, velocity and acceleration, but also by weather and lighting conditions and used agent implementations or driving functions.
	Additionally, two different agent implementation can lead to different scenario outcomes even if the same variables are used.
	Finding the outcome of a scenario can be viewed as a multidimensional optimization problem, since all this has to be considered before the scenario modelling process and can be used to create new concrete scenarios or to train driving functions via reinforcement learning \cite{abeysirigoonawardena2019generating}.
	Two possible and most obvious outcomes of the urban intersection scenario could be \textbf{collision} or \textbf{no collision}. 
	However, the outcome is usually evaluated by pass and fail criteria that are define by the System-under-Test (SuT) and the reason a scenario is used, e.g. validation or verification, reinforcement learning, etc., and can be of a more complex nature than a collision between two actors, e.g. time dependent conflict measures as mentioned by \cite{astarita2019traffic}.
	Another approach to model scenarios was proposed by \citet{majzik2019towards}, where a context model and a situation model define a scenario.
	The context model describes the scene and the environment, and the situation model holds all object information.
	During the execution of a scenario, the situation model is constantly updated with the help of the context model, but a clear sequence of what is happening in a scenario is not supported.
	
	A tool-independent scenario description standard currently under development is ASAM OpenSCENARIO. 
	In March 2020 version 1.0 of this standard was released.
	It works together with ASAM OpenDRIVE and OpenCGR which complement each other to describe all aspects of a traffic scenario: from the road topology to all participating vehicles and their maneuvers \cite{openscenario}. 
	OpenSCENARIO only supports concrete scenarios, but allows parameterization in order to load different parameter sets to realize logical scenarios.
	Additionally, it supports a catalog concept for maneuvers, actions, trajectories among others \cite{openscenario} for a better reusability.
	OpenSCENARIO uses a complex XML syntax, which makes it difficult for humans to understand or even design a new scenario by hand.
	
	The CARLA simulator, a simulator based on the Unreal gaming engine, and its scenario runner already support a textual behavior tree structure based on a python behavior tree language \cite{dosovitskiy2017carla,scenariorunner}. 
	A scenario is modeled without position data or actor type.
	This information is given in an extra configuration file, which contains initial values for several different parameter sets.
	Since not all parameters can be varied in this file, e.g. vehicle velocities, CARLA offers concrete and partial logical scenarios.
	However, as of now CARLA does not support modularity or reusable code, since a new scenario has to be completely designed and written from scratch by an expert.
	
	All of the mentioned scenario description and modeling languages fulfill one ore more of our proposed features, but none fulfills all of them.
	
	\section{Scenario Modeling Language}
	\label{sec:language}
	\begin{figure}[ht]
		\centering
		\includegraphics[width=1\linewidth]{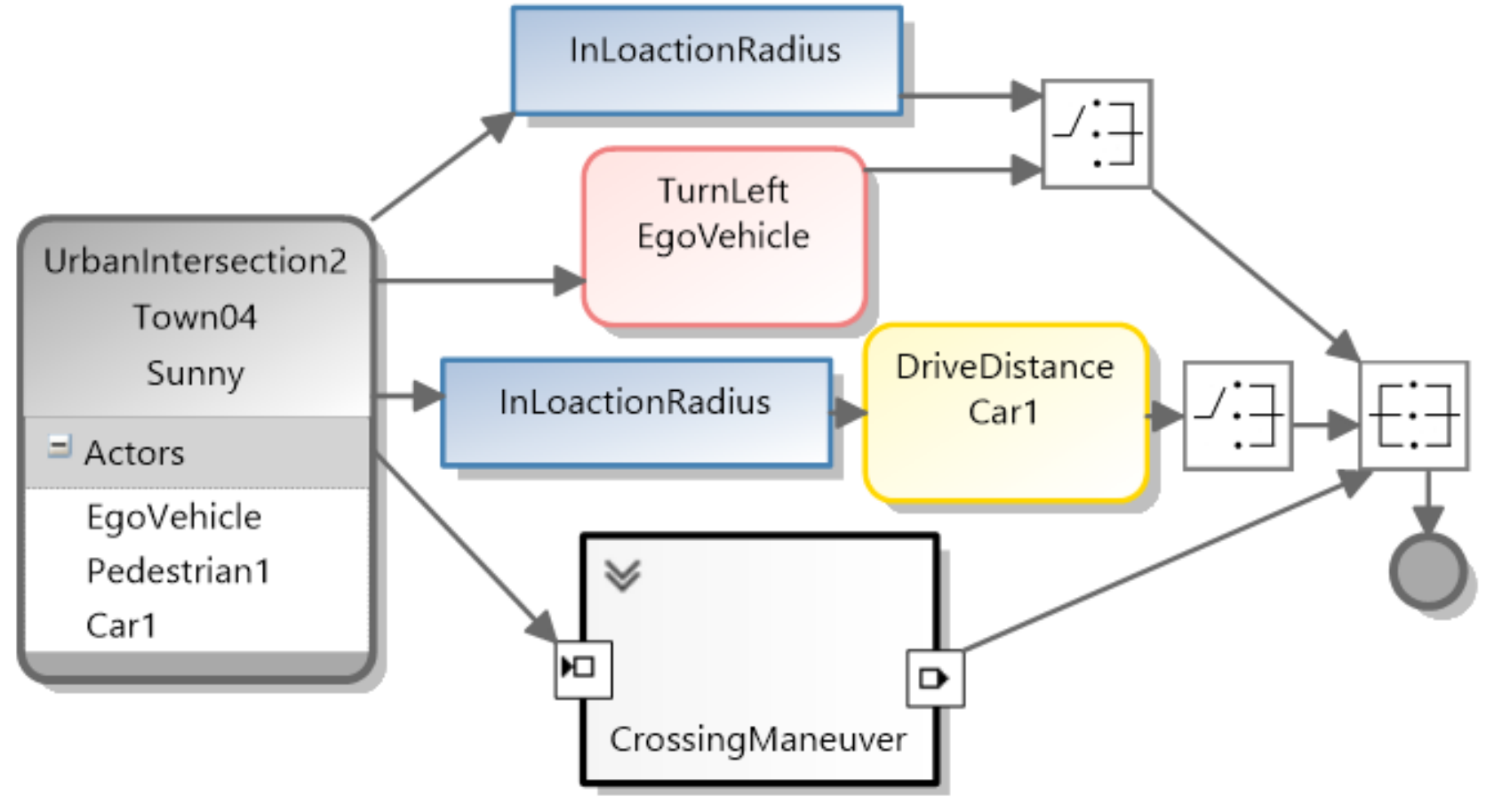}
		\caption{Second urban intersection scenario with ego vehicle, car and pedestrian.}
		\label{fig:second_scen}
	\end{figure}
	Our scenario modeling language (SceML) is explained by using two examples.
	The first example scenario is the already used urban intersection scenario (UIS1) depicted in Fig.~\ref{fig:complete}d), which shows this scenario in our proposed graph notation.
	The start situation and the actors' routes are depicted in Fig.~\ref{fig:complete}a). 
	The ego vehicle (blue car) wants to make a right turn. 
	The bike, driving on a parallel lane to the ego vehicle, crosses the street right after the intersection. 
	Additionally, the scenario contains three synchronization conditions to schedule the actors' actions within an scenario execution:
	\begin{itemize}
		\item \textbf{Sync 1:} Ego vehicle reaches destination around the corner, depicted as the top blue \textit{InLocationRadius} node in Fig.~\ref{fig:complete}d) .
		\item \textbf{Sync 2:} Ego vehicle approaches the bike within a certain radius to signal the bike to start, depicted in the left bottom blue \textit{InVehicleRadius} node.
		\item \textbf{Sync 3:} Bike reaches destination across the road, depicted as the right blue \textit{InLocationRadius} node in Fig.~\ref{fig:complete}d).
	\end{itemize}
	
	\begin{figure}[ht]
		\centering
		\includegraphics[width=\linewidth]{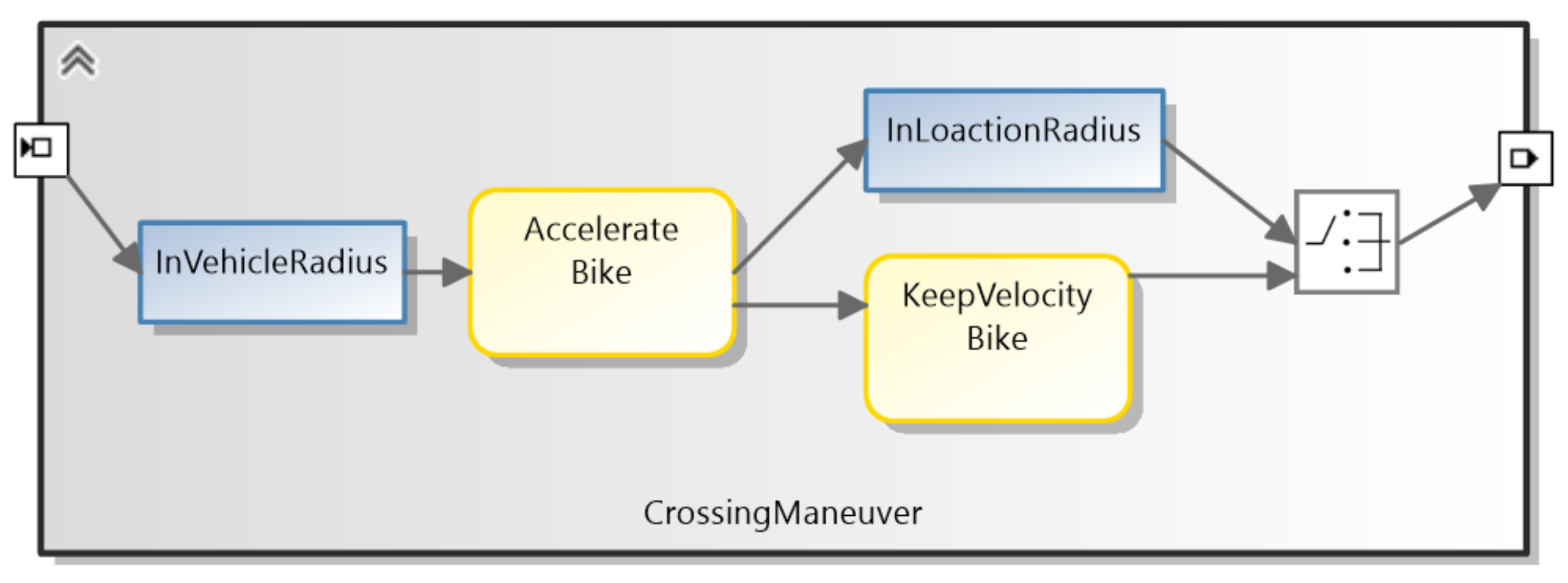}
		\caption{Expanded CrossingManeuver which is shown collapsed in Fig.~\ref{fig:second_scen}. }
		\label{fig:mods}
	\end{figure}
	
	The second example scenario takes place at the same urban intersection and is shown in SceML graph notation in Fig.~\ref{fig:second_scen} (UIS2).
	The ego vehicle starts in the same position as in UIS1, however, instead of a right turn it wants to turn left.
	A second actor is an oncoming car, which simply wants to cross the intersection without turning left or right.
	The last actor in this scenario is a pedestrian crossing the intersection in the same manner as the bike from UIS1 with the difference that it is moving on the left side walk instead of the right.
	UIS2 also has a number of synchronization conditions:
	\begin{itemize}
		\item \textbf{Sync 1:} Ego vehicle reaches destination around the corner, depicted as the top blue \textit{InLocationRadius} node in Fig.~\ref{fig:second_scen}.
		\item \textbf{Sync 2:} Ego vehicle approaches location within the intersection to signal the second car to start, depicted by the blue \textit{InLocationRadius} node set before the DriveDistance nod.
		\item \textbf{Sync 3:} Second car is close to pedestrian to signal the pedestrian to start crossing the street, depicted by first \textit{InVehicleRadius} node in CrossingManeuver shown in Fig.~\ref{fig:mods}.
		\item \textbf{Sync 4:} Pedestrian reaches destination across the road, depicted as the right blue \textit{InLocationRadius} node in Fig.~\ref{fig:mods}.
	\end{itemize}

	\subsection{Notation}
	\label{sec:notation}
	
	\begin{figure}[ht]
		\centering
		\includegraphics[width=\linewidth]{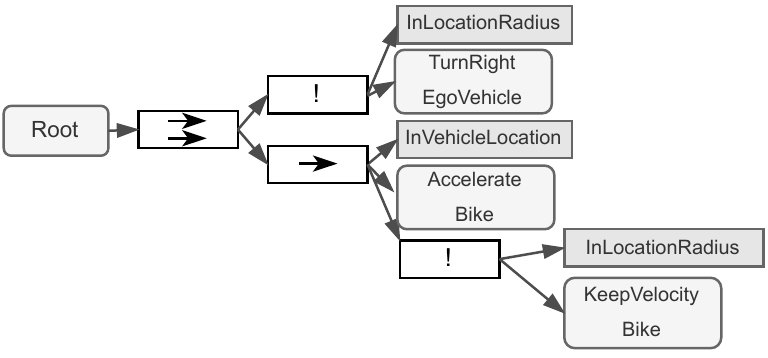}
		\caption{Urban intersection scenario from Fig.~\ref{fig:complete}d) represented in behavior tree notation as it was proposed by \citet{colledanchise2016behavior}.}
		\label{fig:treenot}
	\end{figure}
	The notation of SceML is graph-based and has similarities to reversed behavior trees as they were proposed by \cite{colledanchise2016behavior} and \cite{marcotte2017behavior}, an established way of expressing and structuring artificial intelligence in games. 
	Usually, a behavior tree starts at exactly one root node and has several children, that can either be behavior nodes or sub-trees with different execution policies: sequential execution, displayed by a single arrow node, parallel execution with demanded success on one child (question mark node) or parallel execution with demanded success on all children (double arrow node).
	Fig.~\ref{fig:treenot} shows UIS1 with the aforementioned synchronization conditions as behavior tree.
	
	SceML utilizes the same three execution policies, however, due to human readability depicts them in a different way. 
	Sequential execution is shown by a sequential chain of maneuvers and conditions and, thus, does not need an extra symbol. 
	Parallel executed maneuvers and conditions join in dedicated join nodes: to proceed further at least one path has to be finished or all paths have to be finished.
	
	\begin{table}
		\caption{Node Notation Legend}
		\label{tab:nodenotation}
		\begin{tabular}{l@{}l@{}|l@{}l@{}}
			\toprule
			Node&Description& Node&Description \\
			\midrule
			\begin{minipage}{1cm}
				\includegraphics[height=0.6cm]{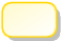}
			\end{minipage} & Longitudinal maneuver &
			\begin{minipage}{1cm} 
				\includegraphics[height=0.57cm]{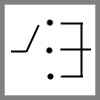}
			\end{minipage} & Join: one finished\\
			\begin{minipage}{1cm}
				\includegraphics[height=0.6cm]{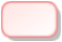}
			\end{minipage} & Lateral maneuver &
			\begin{minipage}{1cm} 
				\includegraphics[height=0.57cm]{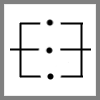}
			\end{minipage} & Join: all finished\\
			\begin{minipage}{1cm}  
				\includegraphics[height=0.35cm]{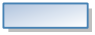}
			\end{minipage} & Condition &
			\begin{minipage}{1cm}
				\includegraphics[height=0.6cm]{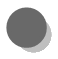}
			\end{minipage} & End node\\    
			
			\bottomrule
		\end{tabular}
	\end{table}
	
	\subsubsection{Root and end node}
	Each scenario has one root node, that holds important scenario information, e.g. name, actor list, etc.
	In addition, each scenario has an end node (see Tab.~\ref{tab:nodenotation}), where all maneuver paths have to end.
	
	\subsubsection{Maneuver and condition nodes}
	The left hand side of Tab.~\ref{tab:nodenotation} shows a list of possible maneuver nodes: longitudinal - movements, parallel to the actor's movement direction (yellow), lateral - movements perpendicular to the actor's movement direction (red). 
	More complex maneuvers, which consist of a combination of longitudinal and lateral movements are depicted in green. 
	However, they are not shown here, since these composite maneuvers are not used in UIS1 and UIS2. 
	
	Condition nodes help to synchronize maneuvers or to make sure certain parts of a graph are only executed when a set of requirements is fulfilled and are colored blue.
	
	\subsubsection{Join nodes}
	Join nodes symbolize the parallel execution of nodes or sequences. 
	A \textit{Join-all-finished}-Node denotes a successful execution of all incoming nodes or sequences is required. 
	A \textit{Join-one-finished}-Node is viewed as successfully executed when one possible path to this node is executed.
	Both join node types are shown in Tab.~\ref{tab:nodenotation}.

	\subsubsection{Modules}
	Modules allow to summarize maneuvers and conditions for various reasons:
	\begin{itemize}
		\item[1.] A combination of basic maneuvers results in a more complex maneuver, e.g., lane change to the left, pass vehicle, lane change to the right. 
		This maneuver sequence can be put in a module frame, which is called \textit{Overtaking Maneuver}.
		Modules allow collapsed versions of a tree and make it easier to understand a scenario.
		\item[2.] Modules can be saved in a library and reused in further scenarios.
	\end{itemize}
	Additionally, modules are allowed to contain sub-modules.
	Fig.~\ref{fig:second_scen} shows a scenario using the \textbf{CrossingManeuver} module, which is the same sequence of actions and conditions the bike in UIS1 is executing.
	A module has in- and output ports in order to manage incoming and outgoing connections.

	\subsection{Language Metamodel}
	\begin{figure}[ht]
		\centering
		\includegraphics[width=1.0\linewidth]{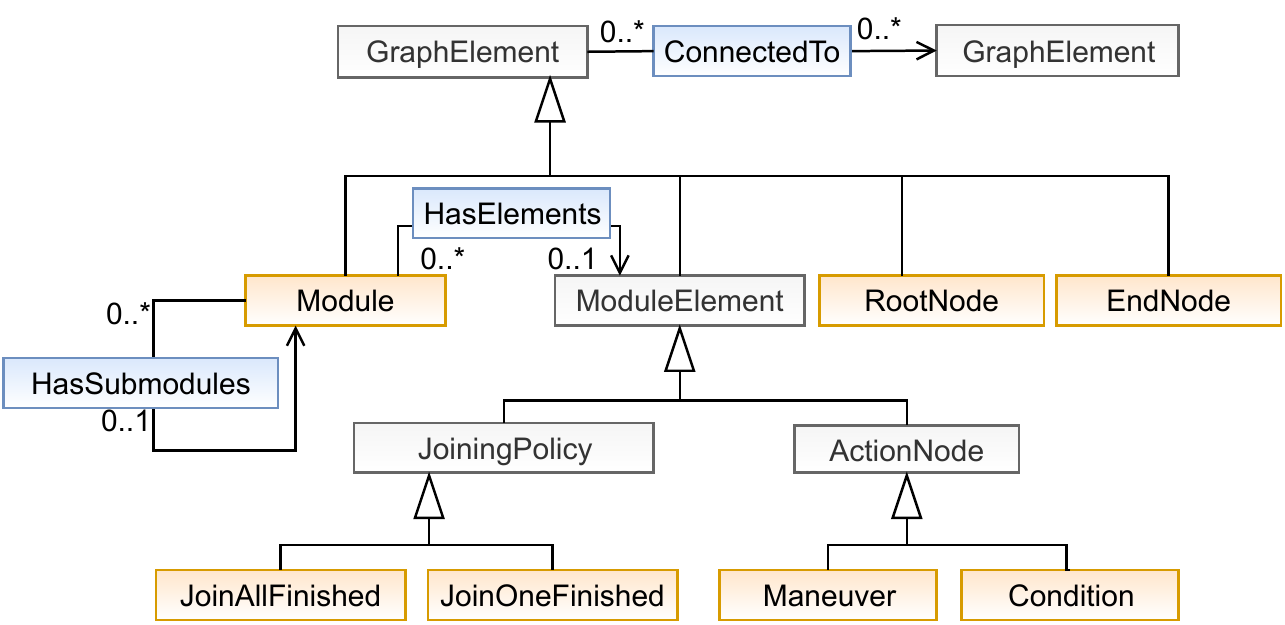}
		\caption{Partial metamodel of SceML's graph structure.}
		\label{fig:metamodel}
	\end{figure}
	
	Fig.~\ref{fig:metamodel} shows a partial metamodel of SceML's graph structure. 
	It mostly follows UML class diagram conventions and contains all important classes. 
	Classes colored in grey are abstract classes, whereas orange classes are not. Blue boxes represent relations between classes.
	\subsubsection{Graph Structure}
	
	The whole scenario description is based on a hierarchical graph structure.
	Each graph node inherits from \textit{GraphElement}, which can be further categorized into \textit{Module}, \textit{ModuleElement}, \textit{RootNode} and \textit{EndNode}.
	The graph's root, called \textit{RootNode}, holds information about actors, their starting positions, environment, scenario name, map name etc. 
	All possible paths have to end in one \textit{EndNode}. 
	If no further information about joining is given, the end node works as a join-all-finished node.
	Every scenario can only contain one root and one end.
	
	A \textit{Module} is a combination of different \textit{ModuleElement}s.
	Only root and end nodes cannot be part of a module, since a module is defined as a complex maneuver.
	\textit{ModuleElement}s are all nodes that can be part of a scenario graph and are either \textit{ActionNode}s or \textit{JoiningPolicy} nodes.
	Action nodes again are divided into \textit{Maneuver} nodes, defining different maneuvers or \textit{Condition} nodes, which impose different runtime conditions on an executed scenario.
	\textit{JoiningPolicy} define the type of a parallel execution: success on all or on one.
	The \textit{ModuleElement} can be viewed as an artificial layer in order to make all tree node types available as parts of a module.
	
	\subsubsection{Modules}
	The main feature of SceML is a modular scenario concept and scenario description, i.e., to build and design scenarios and reusable sub-scenarios. 
	A \textit{Module} can contain none to several sub-modules and an instance of a module can only be part of one parent module. 
	However, different instances of a module can be parts of different parent modules. 
	Additionally, a module can hold none to several module elements, but again the instance of a module element can only be part of one module. 
	For example, a module \textit{m} can contain several instances of \textit{TurnRight}, but each instance of \textit{TurnRight} can only be part of module \textit{m}.
	The relation \textit{HasSubmodules} in Fig.~\ref{fig:metamodel} describes the ability for a module to contain other modules as sub-modules.
	The possible depth for sub-modules within sub-modules is in theory infinite and only limited by computational resources and clarity of a scenario.
	
	\subsubsection{Actors}
	\begin{figure}[ht]
		\centering
		\includegraphics[width=0.65\linewidth]{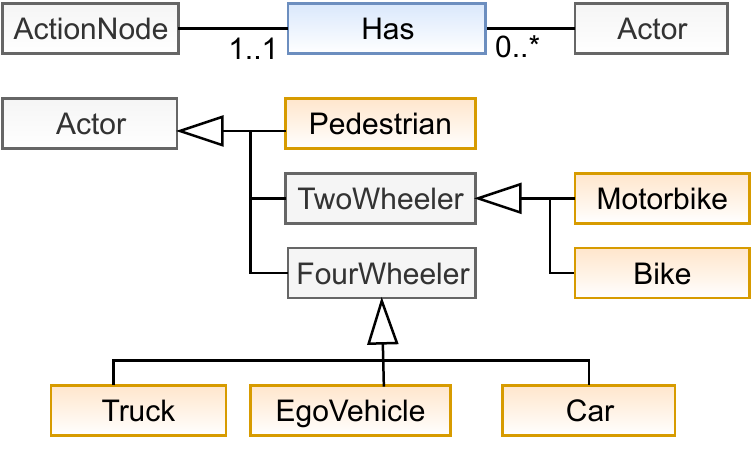}
		\caption{ Metamodel describing actors and their relationship to action nodes. }
		\label{fig:actors}
	\end{figure}
	Actors are divided into three categories: pedestrians, two wheelers and four wheelers.
	The proposed class-based definitions do not claim completeness, but show an intuitive way to structure participant types in an expandable way. 
	Each maneuver or condition has a relation to exactly one actor, as shown in Fig.~\ref{fig:actors}.
	This actor is called reference actor and describes the actor executing a maneuver or in need to fulfill a condition.
	Each actor can be reference actor of none up to several action nodes.
	An action node can also have a relation to a second actor, called target actor.
	Examples for such maneuvers are \textit{Follow} or \textit{Approach}.
	
	\subsection{Relation To Abstraction Levels}
	The abstraction level is set as a parameter in the root node and valid for all elements of a scenario.
	A concrete scenario is modeled by a graph with defined parameters for each variable and if no value is given uses default values defined by the language.
	The goal of a concrete scenario is to have an executable scenario and, therefore, is parsed into OpenSCENARIO file format as described in Sec.~\ref{sec:impl}.
	The modeling of a logical scenario is similar to the concrete scenario, but instead of certain values a range can be given.
	Nonetheless, is also possible to use concrete values if no value range is needed.
	Additionally, a functional scenario is the graph model of a scenario without the necessity of given variable ranges or values, but both can be given if wanted.
	Functional and logical scenarios cannot be used to generate proper OpenSCENARIO files, since the description language demands it to be a concrete scenario.
	
	\subsection{Language Validation}
	\label{sec:validation}
	
	The graph structure and the scenario content need to be validated in order to get an executable scenario.
	Some metamodel decisions are made due to model or graph validity.
	A scenario has to have a one root node and a root node belongs to one model.
	The same is true for end nodes.
	All maneuver and condition nodes have to be guaranteed to be part of a path through the directed graph, that is connected to the end node.
	However, further requirements have to be met to get a valid scenario:
	\begin{itemize}
		\item[1.] The root node cannot have an incoming connection and the end node cannot have an outgoing connection.
		\item[2.] Parameters must be given and fit the scenario abstraction level.
		\item[3.] Join nodes need at least two incoming sequences.
		\item[4.] (optional) Certain maneuvers are not executed simultaneously, e.g. an actor cannot accelerate and decelerate at the same time.
		\item[5.] (optional) Parameters fit to the actor and maneuver, e.g. a pedestrian cannot move with a velocity of 50 $\frac{km}{h}$.
	\end{itemize}
	SceML has yet to be extended with custom validation rules to implement these requirements.

	\subsection{Implementation And Execution}
	\label{sec:impl}
	
	To develop an experimental scenario modeling framework, which allows to model a graphical graph-based scenario representation, we used the Microsoft Visual Studio Modeling SDK \cite{msvsdsl}.
	Fig.~\ref{fig:complete} d) shows a scenario graph modeled with our framework.
	\begin{figure}[ht]
		\centering
		\includegraphics[width=0.7\linewidth]{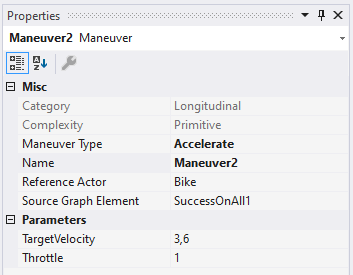}
		\caption{Exemplary property definition for \textit{Accelerate} maneuver executed by the bike.}
		\label{fig:properties}
	\end{figure}
	It is possible to define further information for each node used in a scenario than what is shown in UIS1 and UIS2 graph notation of Fig.~\ref{fig:complete}d) and Fig.~\ref{fig:second_scen}.
	Fig.~\ref{fig:properties} presents a properties menu for the \textit{Accelerate} maneuver assigned to the bike or pedestrian in the CrossingManeuver. 
	It is possible to customize all given parameters under \textbf{Parameters}, e.g. target velocity and throttle for acceleration. 
	\textbf{Misc} describes general maneuver settings, such as name or maneuver type, but also maneuver dependent information that cannot be changed, e.g. maneuver category and complexity level. 
	In order to have a tool compliant scenario file, we build a custom OpenSCENARIO 1.0 parser to parse the graph structure into a valid OpenSCENARIO file.
	
	We used the CARLA simulator version 0.9.9 \cite{dosovitskiy2017carla} and its scenario runner expansion \cite{scenariorunner} to execute UIS1 and UIS2. 
	CARLA's scenario runner partially supports OpenSCENARIO 1.0 since version 0.9.9 as possible scenario input.
	Fig.~\ref{fig:complete}e) shows two screenshots from the scenario execution. 
	The upper image was taken directly after scenario initiation and the lower image during execution time, while both actors drive into the intersection but do not collide. 
	For both scenarios the option exists to change variables in order to create collisions or otherwise critical situations.
	UIS1 could end in a collision between ego vehicle and bike, whereas UIS2 can end without a collision, a collision between ego vehicle and car, or a collision between ego vehicle and pedestrian.
	
	In theory it is possible to support different data types, e.g. world or road coordinates, for actor position and enable different presentation modi in maneuver and condition nodes.
	However, we did not implement it at this stage, since CARLA in contrast to OpenSCENARIO is not supporting it at this time.

	\section{Conclusion And Future Work}
	\label{sec:conclusion}
	This work extends previous scenario modeling approaches by combining a graphical approach with the possibility to model different abstraction levels.
	In detail, a concept for a graphical scenario modeling framework was presented. It uses a graph structure to represent scenarios for the development and testing process of autonomous vehicles.
	In the present state, it allows to model scenarios in different abstraction levels. 
	A functional scenario can be modeled by using node representations without further information about given maneuvers and variables.
	The properties of all nodes can be filled with certain values, in case an executable concrete scenario is needed or with value ranges if a logical scenario is desired.
	Our approach supports modularity and reusable sub-scenarios, that can be stored in a library. 
	A custom OpenSCENARIO parser was used to store concrete scenarios and enable the CARLA simulator to load a scenario.
	
	Our modeling approach can be used to create concrete scenarios for test cases used in the validation process, but further holds the possibility to only create logical scenarios for a later exploration to find critical and interesting concrete scenarios, since the outcome is not defined before execution.
	Hence, this approach can be used to find new scenarios to build and complete data bases in particular by providing the possibility to combine human and computational efforts.
	Finally, SceML could be used to inspect and modify automatically generated and extracted scenarios from recorded data sets.
	
	As a next step we plan to implement missing validation requirements and to combine our modeling framework with a scenario evaluation framework, in order to only model a logical scenario which is then used to find all interesting concrete scenarios, i.e. collisions, near collisions and other critical conflict situations.

	\bibliographystyle{unsrtnat}
	\bibliography{papermodlang}  

	
	
	
	
	
\end{document}